\documentclass[sn-mathphys-num]{sn-jnl}

\usepackage{graphicx}%
\usepackage{multirow}%
\usepackage{amsmath,amssymb,amsfonts}%
\usepackage{amsthm}%
\usepackage{mathrsfs}%
\usepackage[title]{appendix}%
\usepackage{xcolor}%
\usepackage{textcomp}%
\usepackage{manyfoot}%
\usepackage{booktabs}%
\usepackage{algorithm}%
\usepackage{algorithmicx}%
\usepackage{algpseudocode}%
\usepackage{listings}%

\theoremstyle{thmstyleone}%
\theoremstyle{thmstyletwo}%

\theoremstyle{thmstylethree}%

\begin{document}

\title[Investigation of the FCNC couplings between the top quark and the photon in $\gamma\gamma$ and $\gamma^{*}\gamma^{*}$ collisions at the CLIC]{Investigation of the FCNC couplings between the top quark and the photon in $\gamma\gamma$ and $\gamma^{*}\gamma^{*}$ collisions at the CLIC}


\author*[1]{\fnm{E.} \sur{Alici}}\email{edaalici@beun.edu.tr}

\affil*[1]{\orgdiv{Department of Physics}, \orgname{Zonguldak Bulent Ecevit University}, \orgaddress{ \city{Zonguldak}, \postcode{67100}, \country{Turkey}}}


\abstract{The investigation of flavour-changing neutral current transitions(FCNC) involving the top quark are only possible with new theoretical frameworks that extend the Standard Model(SM), given that these transitions are almost entirely suppressed within the SM. In this regard, examining these transitions with extended theories beyond the SM may provide significant insights into future experiments. To this end, we present a phenomological study about the anomalous FCNC transitions via $tq\gamma$ couplings with effective field theory. Here, ${\gamma \gamma}{\rightarrow} {\textit{l}} {\nu}_\textit{l} b \bar{q} $ , ${e^{+}} {e^{-}} {\rightarrow} {e^{+}} {\gamma}^{} {\gamma}^{} {\rightarrow} {e^{+}} \textit{l}{} {\nu_\textit{ l}} b \bar{q} {e^{-}} $, $\gamma \gamma {\rightarrow} j \bar{j} b \bar{q} $ and ${e^{+}} {e ^{-}} {\rightarrow} {e^{+}} {\gamma}^{} {\gamma}^{} {e^{-}} {\rightarrow}{e^{+}} j \bar{j} b \bar{q} {e^{-}} $ processes are investigated for the CLIC at $\sqrt{s} =1.5$TeV and 3 TeV. Thus, we have calculated the constraints at $95{\%}$ confidence level on the $BR(t\rightarrow q\gamma)$. As a result, the obtained constraints are at least one order of magnitude better than the available experimental ones.}




\maketitle

\section{Introduction}\label{sec1}

The Standard Model (SM) is a profoundly insightful theoretical framework that elucidates the behaviour and interactions of subatomic particles. It is a validated and powerful theory of particle physics that has proven its accuracy and reliability through the experimental discovery of all subatomic particles foreseen in the SM, including the Higgs boson \cite{hh,hh1}. However, it has some important deficiencies and therefore, it needs to be handled as an fundamental model of an extended theory. In this regard, the scientific community is engaged in the pursuit of developing novel extended model theories, which are referred to as "beyond the Standard Model" (BSM) and they have so far proposed various BSM theories such as standard model effective lagrangian theory (SMEFT), Little Higgs, Supersymmetry, Extra Dimensions, etc. Top quark interactions have an significant mission in the study of extended theories due to their enormous masses, which are almost the same amount as the electroweak scale, and offer a distinctive avenue for exploration and understanding to these theories \cite{1}. One of the striking characteristics of the main modes for producing top quark is that the flavour-changing neutral current (FCNC) transitions are extremely ineffective owing to the Glashow-Iliopoulos-Maiani (GIM) mechanism. Furthermore, it is widely acknowledged that these interactions are prohibited at the tree level and significantly suppressed at the loop level within the GIM mechanism. As it was furthermore established, the total number of these transitions in the SM is well below the experimentally detectable level, for example, the expected branching ratio for $t\rightarrow q\gamma$ $\textit{(q=u,c)}$ decays is about $10^{-14}$ \cite{2,3}. On the other hand, in BSM theories the GIM mechanism is usually not as influential as in the SM and hence, it is possible to observe FCNC effects in the BSM model in high energy particle colliders \cite{4,5,6,7,8,9,10,11,12}. Therefore, phenomenological research on these unusual top quark decay channels will be a proof for extended theories \cite{14,15,16,17,18,19,20,21,22,23,24,25,26,27,28,29,30,31,32,33,34,35,36,37,ch1,ch2}. In the literature so far, several experimental studies have been reported on FCNC couplings of the top quark to the photon and $u$ or $c$ quark and   constraints on the branching fractions of top quark FCNC transitions were acquired by different collaboration groups \cite{38,39,40,41,42,43,44,45,46,47,48,49,50,51,cms,atlas}. In this regard, to date, the strongest constraints on anomalous FCNC $tq\gamma$ interactions have been obtained by the ATLAS and CMS collaborations at the LHC \cite{cms,atlas}. Accordingly, the CMS collaboration reported in their recent research at 2023 that experimental upper limits on the branching ratios (BR) have been achieved: $BR(t\rightarrow u\gamma)<0.95\times10^{-5}$ and $BR(t\rightarrow c\gamma)<1.51\times10^{-5}$ \cite {cms}. In an different experimental research conducted in 2023, ATLAS group has achieved limits on the branching fraction via left-handed(right handed)tq$\gamma$ couplings as $BR(t\rightarrow u\gamma)<0.85\times10^{-5}(1.2\times10^{-5})$ and $BR(t\rightarrow c\gamma)<4.2\times10^{-5}(4.5\times10^{-5})$ \cite{atlas}. The findings yielded by the two research teams are similar each other and report the most stringent experimental constraints up to date. Notwithstanding these recent experimental studies and numerous previous experimental studies, no evidence of anomalous FCNC transitions in the top quark has yet been detected at the center of mass energies of the available colliders. It is evident that this situation can only be overcome by possible future new particle accelerators that can operate at higher center of mass energies than the current ones.
In this respect, it is proposed by different collaborations to construct new particle accelerators for the post-LHC area. These accelerators will be two types which are circular collider and linear collider. The Linear ones of them, the Compact Linear Collider (CLIC), is multi-TeV scale electron-positron collider with high luminosity that relied on new two-beam acceleration systems.  Due to their main advantages such as low cost and clean electromagnetic interaction channels, $\textit{e}^{+}\textit{e}^{-}$ linear colliders are quite remarkable than other collider type. On the other hand, according to conceptual design reports published by CERN in 2012 and revised in 2016, CLIC is envisaged to be operated in phases \cite{clic1,clic2,clic3}. In these phases, it is planned to increase the center of mass energy gradually. Firstly, operation at center of mass energy of 380 GeV will be activated. This phase gives opportunities to the precision examination of Higgs boson and top quark properties accurately. Secondly, center of mass energy will be upgraded to 1.5 TeV. With increased center of mass energy, the CLIC can be enable to discovery significant evidence about physics beyond the SM. Lastly, the collider will be operated with center of mass energy of 3 TeV. This phase with ultimate high energy can open up new horizons for unknown physics beyond the SM, as in the second phase. It is also proposed that each phase will be running for almost four to five years. This situation will make it possible to achieve very high luminosities in the range from $2500 fb^{-1}$ to $5000 fb^{-1}$.
The CLIC has also photon-photon collision options. These collisions can take place of in two different approaches. In one of these approaches, the photons$(\gamma)$ , which are named Compton backscattered (CB) photons, occur with Compton backscattering  of laser light from relativistic electrons. With regard to obtaining that high energy real photons, it will be established additional compton ring and linac to the main collider. This approach offers a promising method for achieving photons, although it also presents a potential drawback. As it requires new equipment, and this means also another cost. In the other alternative approach, the photons$({\gamma}^{*})$, which are called quasi-real photons, emerge from the spontaneous radiation of the main electron or positron beam. The electron or positron beam moving in the main tunnel loses some of its transverse momentum and emitting of photons scattered from the direction of movement at a very small angle. These emitting photons are in accordance with the Weizsacker-Williams approximation (WWA)\cite{65,66,67}.
In light of the above-mentioned important literature, in this study, we performed sensitivity analyses on the branching ratios of FCNC $tq\gamma$ transitions via WW and CB photon interactions using the SMEFT framework for CLIC at 1.5 and 3 TeV center of mass energies. The findings offer promising insights into the existence of FCNC transitions.

\section{THEORETICAL FORMALISM}

An effective field theory (EFT) of a physical system is a theoretical framework used to describe its low-energy behaviour. The Standard Model Effective Field Theory (SMEFT), an effective field theory combined with the SM, allow us to investigate the SM at energies lower than those where it is expected to break down. It is also well established that SMEFT is particularly useful in addressing issues related to the quantum corrections and higher-dimensional operators that may arise in the SM. It, furthermore, make possible physicists to explore potential new physics beyond the Standard Model by analysing the deviations from the SM predictions in experimental data. This is accomplished by expanding the SM Lagrangian in the energy scale divided by powers of new physics scale, which helps in organizing the higher-dimensional operators. In this context, the effective Lagrangian, which incorporates additional terms derived from the SM Lagrangian, is defined as follows \cite{70,71},

\begin{eqnarray}
{\mathcal{L}}={\mathcal{L}}^{(4)}_{\textit{SM}}+\sum_{n=1}\frac{1}{{\Lambda}^n}{\mathcal{L}^{(n+4)}}+h.c..
\end{eqnarray}

where ${\mathcal{L}}^{(\textit{n+4})}$ contains  operators of dimension ${\textit{n+4}}$ produced by the SM  fields, ${\mathcal{L}}_{\textit{SM}}$ is defined the SM Lagrangian and ${\Lambda}$ is the energy scale of new physics theories. In the presented study, it has been taken into the second order terms of the new physics which originated from the dimension six operators describing FCNC couplings among the photon ($\gamma$), top quark ($t$) and q quark ($u$ or $c$). Finally, The SM Lagrangian could be readily expanded with the FCNC couplings in the vertex of tq${\gamma}$ as detailed below \cite{70,71},
	
\begin{eqnarray}
{\mathcal{L}}=\frac{g_{e}}{2m_{t}}\sum_{q=u,c}\bar{q}{\sigma_{\mu\nu} }({\lambda}_{qt}^{R}P_{R}+{\lambda}_{qt}^{L}P_{L})t A^{{\mu}{\nu}}+h.c
\end{eqnarray}
Here, $g_{e}$ corresponds the electromagnetic coupling constant and $\sigma_{\mu\nu}=[\gamma_{\mu} \gamma_{\nu}]/2  $. $ {\lambda}_{qt}^{R(L)}$ corresponds dimensionless real parameters which represent the anomalous couplings constant. In the present work, for the sake of simplicity, it is assumed that there is no specific chirality associated with the FCNC interaction vertices, i.e. $ {\lambda}_{q}={\lambda}_{qt}^{L}={\lambda}_{qt}^{R}$. Besides, in the study, and $ \lambda_{uc}$ and $ \lambda_{ut}$ has been assumed as equal. Using Eq(2), decay width of the anomalous tq${\gamma}$ couplings could be readily calculated as ${\Gamma(t{\rightarrow}q\gamma)}={\frac{\alpha}{2}}{{\lambda}_{q}^2 } {m_{t}}={0.6724}{\lambda_{q}^2} $.  Here, $m_t=173.0$ GeV and the fine structure constant is denoted by  $\alpha$, is related to  ge as $\alpha=(g_{e}e)^2/(4\pi)=\frac{1}{132.1}$. On the other hand, it also is well-established that the branching ratio for the anomalous coupling interaction is defined following, $
  BR(t\rightarrow q\gamma)=\frac{\Gamma(t\rightarrow q \gamma)}{\Gamma(t\rightarrow Total)}$. In the SM, the dominant decay mode of the top quark sector is the channel in which a b quark $(t\rightarrow W b)$ and a W boson are occured, and this channel is roughly equal to the total decay rate of the top quark ${\Gamma(t\rightarrow Total)}$, which is about 1.47 GeV. In light of the aforementioned inputs, the branching ratio of anomalous couplings can be expressed as a fraction of the $t\rightarrow q\gamma$ ($q=u,c$) decay rate relative to the $t\rightarrow W b$ decay rate. After all, it can be simply get as $Br(t\rightarrow q\gamma)={0.4574{\lambda_{q}^2}}$.
\section{Cross Sections} 
In this paper, we have probed ${\gamma \gamma}{\rightarrow} {\textit{l}} {\nu}_\textit{l}  b  \bar{q} $ ,  ${e^{+}} {e^{-}} {\rightarrow} {e^{+}} {\gamma}^{*} {\gamma}^{*} {e^{-}} {\rightarrow}  {e^{+}} \textit{l} {\nu}_\textit{l}  b  \bar{q}  {e^{-}} $ $(\textit{l} = \textit{e},{\mu}$ and  ${\nu}_\textit{l}={\nu}_\textit{e},{\nu}_{\mu})$, $\gamma \gamma {\rightarrow} j \bar{j}  b  \bar{q}  $ and  ${e^{+}} {e^{-}} {\rightarrow} {e^{+}} {\gamma}^{*} {\gamma}^{*} {e^{-}} {\rightarrow}  {e^{+}} j \bar{j} b \bar{q}  {e^{-}} $ ,(j= u,d,b,c,s and $\bar{j}=\bar{u},\bar{d},\bar{b},\bar{c},\bar{s}$) processes via the SMEFT. In the presented tables and figures,  $\gamma \gamma {\rightarrow}{\textit{l}} {\nu}_\textit{l}  b  \bar{q} $ ,  ${e^{+}} {e^{-}} {\rightarrow} {e^{+}} {\gamma}^{*} {\gamma}^{*} {e^{-}}{\rightarrow}  {e^{+}} \textit{l} {\nu}_\textit{l}  b  \bar{q}  {e^{-}} $ processes are named as leptonic channel and  $\gamma \gamma {\rightarrow} j \bar{j}  b  \bar{q}  $, ${e^{+}} {e^{-}} {\rightarrow} {e^{+}} {\gamma}^{*} {\gamma}^{*} {e^{-}}   {\rightarrow}  {e^{+}} j \bar{j} b \bar{q}  {e^{-}} $ as hadronic channel . We have carried out generation of the relevant processes using the Monte Carlo simulation program MadGraph5$\_$aMC@NLO \cite{73}. An expansion of the SM lagrangian definition within the FeynRules package in the MadGraph \cite{72} was achieved by incorporating a Universal FeynRules Output(UFO) module \cite{75} encompassing the SMEFT lagrangian terms \cite{76}.  In this module, the SM matrix has been expressed in a diagonal form. Nevertheless, in the processes under consideration, it is necessary to permit mixing between the first and second generation of quarks. Accordingly, we enclosed the CKM matrix elements into the program.

In the analysis, we accept background process is ${\gamma \gamma}({\gamma}^{*} {\gamma}^{*}){\rightarrow} Wb\bar{q},WW,\textit{l}{\nu}_\textit{l}(jj)b\bar{q}$. To separate signal from background, we apply stringent kinematic selection criteria on generated events as;
\begin{eqnarray}
p_{t}^{\textit{l},b,\bar{q},j ,\bar{j}}> 20 ,|\eta |^{\textit{l},b,\bar{q},j ,\bar{j}}<2.5
\end{eqnarray}
where, $ p_{t}$ is the transverse momentum of all detected particles. This minimum threshold for transverse momentum significantly reducing soft events that are unlikely to originate from high-mass top quark decays. $|\eta |$ is the pseudorapidity of relevant particles and this cut confines particle detection to central detector regions where detector efficiency and resolution are optimal. It eliminates particles scattered at wide angles, typical of background processes. These cuts were specifically optimized for photon-photon collision events to ensure that only particles within high-energy ranges relevant to FCNC interactions were analyzed, suppressing lower-energy backgrounds from electroweak sources.All the other cuts in program have been accepted as in default settings. 

In the study, The CTEQ6L1 parton distribution function, which includes the WW photon distribution function, was used to perform the analyses. The distribution function of the WW photons in the $ {\gamma}^{*} {\gamma}^{*} $processes is defined as follows, \cite{65,66,67,68},
\begin{eqnarray}
{f_{{\gamma}^*}(x)=\frac{\alpha}{\pi E_{e}}\{[\frac{1-x+x^{2}}{x}]log(\frac{Q_{max} ^ {2}}{Q_{min} ^ {2}})-\frac{m_{e}^{2}x}{Q_{min} ^ {2}}(1-\frac{Q_{max} ^ {2}}{Q_{min} ^ {2}})-\frac{1}{x}[1-\frac{x}{2}]^{2}log(\frac{x^{2}E_{e}^{2}+Q_{max} ^ {2}}{x^{2}E_{e}^{2}+Q_{min} ^ {2}})\}}
\end{eqnarray}
where $x=\frac{E_{{\gamma}^*}}{E_{e}}$. Here, while $Q_{max}^{2}$, maximum photon virtuality, is chosen as  2GeV$^{2}$ and  the minimum photon virtuality ($Q_{min}^{2}$) is described as below

\begin{eqnarray}
{Q_{min} ^ {2}=\frac{m_{e}^{2}x}{1-x}}
\end{eqnarray}

However, the CB spectrum is not defined in the program. For this reason, we added the spectrum of compton backscattered photon distribution manually into the CTEQ6L1.
For the $ {\gamma}{\gamma} $ processes,  the spectrum of CB
photons is defined by \cite{69,70},
\begin{eqnarray}
f_{\gamma}(y)=\frac{1}{g(\zeta)}(1-y+\frac{1}{1-y}-\frac{4y}{\zeta(1-y)}+\frac{4y^2}{{\zeta}^2(1-y)^2})
\end{eqnarray}
here,
\begin{eqnarray}
g(\zeta)=(1-\frac{4}{\zeta}-\frac{8}{{\zeta}^2})log(\zeta+1)+\frac{1}{2}+\frac{8}{{\zeta}}-\frac{1}{2(\zeta+1)^2}
\end{eqnarray}
 Here, $y=\frac{E_{\gamma}}{E_{e}}$ is the fraction of electron energy carried away by the scattered photon. Here, $E_\gamma$ is the energy of the backscattered photon and $E_e$ is initial energy of the electron beam before Compton backscattering. On the other hand, $ \zeta$  equals $\frac{4E_{0}E_{e}}{m_{e}^2}$ and here, $E_0$ is energy of the incoming laser photon. When the value of $ \zeta $ is of 4.8, the $ y$ reaches the maximum value of 0.83 ($ y_{max} =\frac{\zeta}{1+{\zeta}}$). This scenario corresponds with the time period during which the photon conversion efficacy experiences a significant decline due to the production of $e^{+}e^{-}$ pairs from laser photons and photon backscattering.
Consequently, the cross section for the processes taking place in the photon collider can be calculated by integrating the subprocess cross section with the photon luminosity at an $\textit{e}^{+}\textit{e}^{-}$ linear collider. This is performed as follows:

    \begin{eqnarray}
\sigma=\int f_{{\gamma},({{\gamma}^*})}(y)f_{{\gamma},({{\gamma}^*})}(y)d \hat{\sigma} d{E_{{\gamma},({{\gamma}^*})1}}d{E_{{\gamma},({{\gamma}^*})2}}   
  \end{eqnarray}

Using these two types of photon distribution functions mentioned above, the dependence of the cross section on the anomalous coupling constant for leptonic and hadronic decay channels was calculated using two different center of mass energies (1.5 TeV and 3 TeV) for WW photons and CB photons. Accordingly, Figures 1 and 2 represent the calculations performed for 1.5 TeV and 3 TeV center of mass energies of leptonic channel processes, respectively, while Figures 3 and 4 correspond the hadronic channel ones. When whole figures are evaluated together, it is clearly seen that the cross section values calculated for hadronic processes are higher than leptonic processes ones. Moreover, it is clear that the calculated cross section values for CB photon interactions are considerably higher than those for WW photon interactions. This finding may indicate that CB photon interactions are more promising to investigate new physics contributions.

\begin{figure}
  \centering
  \includegraphics[width=10.cm]{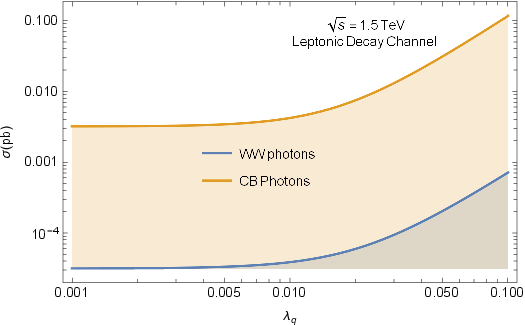}
   \caption{The total cross sections of the processes $ \gamma \gamma   {\rightarrow}   \textit{l} {\nu}_\textit{l}  b  \bar{q}   $ and  ${e^{+}} {e^{-}} \rightarrow {e^{+}} {\gamma}^{*} {\gamma}^{*} {e^{-}} {\rightarrow}  {e^{+}} \textit{l} {\nu}_\textit{l}  b  \bar{q}  {e^{-}} $ as a function of the anomalous ${\lambda}_{q}$ coupling for center-of mass energy of $\sqrt{s}=1.5  TeV$ at the CLIC.}\label{FIG9.}
\end{figure}

\begin{figure}
  \centering
  \includegraphics[width=10.cm]{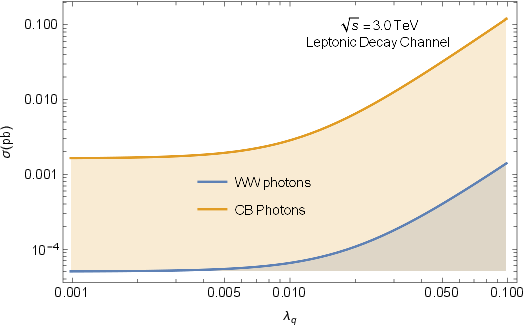}
   \caption{The total cross sections of the processes $\gamma \gamma   {\rightarrow}W b \bar{q} {\rightarrow} \textit{l} {\nu}_\textit{l}  b  \bar{q}   $ and  ${e^{+}} {e^{-}} \rightarrow {e^{+}} {\gamma}^{*} {\gamma}^{*}{\rightarrow} {e^{-}} {e^{+}} W b \bar{q} {e^{-}}{\rightarrow} {e^{+}}\textit{l} {\nu}_\textit{l} b \bar{q}  {e^{-}} $ as a function of the anomalous ${\lambda}_{q}$ coupling for center-of mass energy of $\sqrt{s}=3.0TeV$ at the CLIC.}\label{FIG9.}
\end{figure}

\begin{figure}
  \centering
  \includegraphics[width=10.cm]{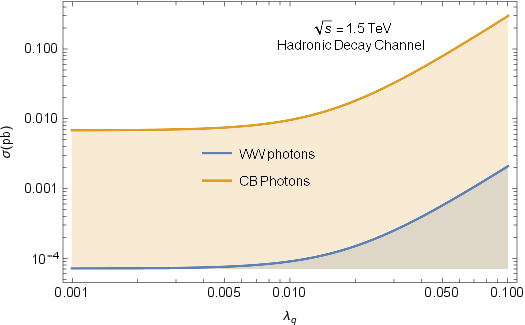}
   \caption{The total cross sections of the processes $ \gamma \gamma   {\rightarrow}W b \bar{q} {\rightarrow}   j \bar{j}  b  \bar{q}   $ and  ${e^{+}} {e^{-}} \rightarrow {e^{+}} {\gamma}^{*} {\gamma}^{*} {e^{-}}  {\rightarrow}  {e^{+}} j \bar{j} b \bar{q}  {e^{-}} $ as a function of the anomalous ${\lambda}_{q}$ coupling for center-of mass energy of $\sqrt{s}=1.5TeV$ at the CLIC.}\label{FIG9.}
\end{figure}
\begin{figure}
  \centering
  \includegraphics[width=10.cm]{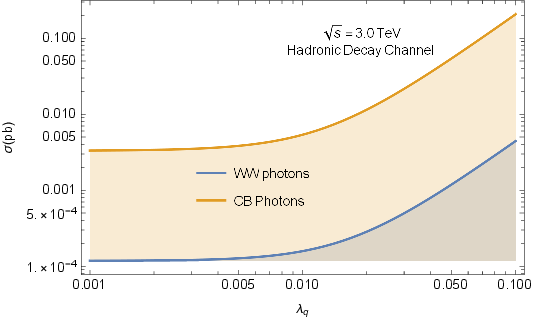}
   \caption{The total cross sections of the processes $\gamma \gamma  {\rightarrow} W b \bar{q} {\rightarrow}  j \bar{j}  b  \bar{q}   $ and  ${e^{+}} {e^{-}} \rightarrow {e^{+}} {\gamma}^{*} {\gamma}^{*} {e^{-}}  {\rightarrow}  {e^{+}} j \bar{j} b \bar{q}  {e^{-}} $ as a function of the anomalous ${\lambda}_{q}$ coupling for center-of mass energy of $\sqrt{s}=3.0TeV$ at the CLIC.}\label{FIG9.}
\end{figure}

\section{Sensitivity Analysis}
In examining the sensitivity of anomalous ${\lambda}_q$ couplings, we have employed the $\chi^{2}$ criterion through the ${\gamma \gamma}{\rightarrow} {\textit{l}} {\nu}_\textit{l}  b  \bar{q} $ ,  ${e^{+}} {e^{-}} {\rightarrow} {e^{+}} {\gamma}^{*} {\gamma}^{*} {\rightarrow}  {e^{+}} \textit{l}{} {\nu}_\textit{l}  b  \bar{q}  {e^{-}} $, $\gamma \gamma {\rightarrow} j \bar{j}  b  \bar{q}  $ and  ${e^{+}} {e^{-}} {\rightarrow} {e^{+}} {\gamma}^{*} {\gamma}^{*} {e^{-}} {\rightarrow}{e^{+}} j \bar{j} b \bar{q}  {e^{-}} $ processes. The $\chi^{2}$ criterion is a statistical measure used to evaluate the difference between the predicted and observed values, allowing for the assessment of how well a theoretical model aligns with experimental data. In this study, the $\chi^{2}$ criterion quantifies the deviation between the Standard Model cross-section ($\sigma_{SM}$) and the total cross-section ($\sigma_{total}$)  which includes contributions from both the Standard Model and new physics. It is defined as follows:
	\begin{eqnarray}
	{\chi^{2}=(\frac{\sigma_{sm}-\sigma_{total}}{\sigma_{sm}
{\sqrt{{\delta}^2_{sys}+{\delta}^2_{stat}}}
})^{2}}
	\end{eqnarray}

where,  $ {\delta}_{stat}=\frac{1}{\sqrt{N}} $ shows the statistical error and $ {\delta}_{sys}$ corresponds  systematic uncertainty. In $ {\delta}_{stat}$, N is identified by the formula of $N=L_{int}\times b_{tag} \times \sigma_{SM}$ where $b_{tag}$ symbolizes tagging efficiency and it is chosen $0.7$ \cite{bquark1,bquark2}.

For sensitivity analysis, constraints on the anomalous coupling parameter for systematic uncertainties of $0\%$,$3\%$,$5\%$ have been obtained. In Tables I and II, we represented limits at center of mass energy of 1.5 TeV for integrated luminosities 100,500,1000,1500 and 2500 $fb^{-1}$ . While Table I has been listed results related ${e^{+}} {e^{-}} {\rightarrow} {e^{+}} {\gamma}^{*} {\gamma}^{*} {e^{-}} {\rightarrow}  {e^{+}} \textit{l}{} {\nu}_\textit{l}  b  \bar{q}  {e^{-}} $ and ${e^{+}} {e^{-}} {\rightarrow} {e^{+}} {\gamma}^{*} {\gamma}^{*} {e^{-}} {\rightarrow} {e^{+}} j \bar{j} b \bar{q}  {e^{-}} $ processes, Table II shows data about ${\gamma \gamma}{\rightarrow} {\textit{l}} {\nu}_\textit{l}  b  \bar{q} $ and $\gamma \gamma {\rightarrow} j \bar{j}  b  \bar{q}  $ processes.
Tables III and IV present the same limits as in Tables I and II, but for a center of mass energy of 3.0 TeV and integrated luminosities of 100, 1000, 3000, 4000, and 5000$fb^{-1}$. The obtained results show that the increasing systematic error values worsened the obtained limits for the anomalous coupling constant.

\begin{table}
\caption{95\% C.L. bounds on the anomalous $\lambda_{tq\gamma}$ coupling for center of mass energy $\sqrt{s}=1.5$ TeV through the processes ${e^{+}} {e^{-}} {\rightarrow} {e^{+}} {\gamma}^{*} {\gamma}^{*} {e^{-}}{\rightarrow}  {e^{+}} j \bar{j} b \bar{q}  {e^{-}} $ and ${e^{+}} {e^{-}} \rightarrow {e^{+}} {\gamma}^{*} {\gamma}^{*} {e^{-}}{\rightarrow}  {e^{+}}\textit{l} {\nu}_\textit{l} b \bar{q}  {e^{-}}$.It is taken under consideration, systematic error as $0\%$,$3\%$,$5\%$ and integrated luminosities as 100,500,1000,1500,2500 $fb^{-1}$.}
\begin{tabular}{cccccc}
\hline
\multicolumn{6}{c}{$\sqrt{s}=1.5 $ TeV} \\
\hline
\multicolumn{2}{c}{} & \multicolumn{2}{c}{Hadronic Channel} & \multicolumn{2}{c}{Leptonic Channel} \\
\hline
${\cal L} \, (fb^{-1})$  & \hspace{0.5cm} $ \delta_{sys}$ \hspace{0.5cm}  &
\hspace{1.5cm} $\lambda_{q}$ \hspace{1.5cm} &&
\hspace{1.5cm} $\lambda_{q}$ \hspace{1.5cm} & \\
\hline\hline
100  &  $0\%$   &0.0178421   &&0.0242639&   \\
100  &  $3\%$   &0.0178618   &&0.0242762&   \\
100  &  $5\%$   &0.0178968   &&0.0242979&   \\
\hline\hline
500  &  $0\%$   &0.0120373   &&0.0160745&   \\
500  &  $3\%$   &0.0121029   &&0.0161154&   \\
500  &  $5\%$   &0.0122168   &&0.0161873&   \\
\hline\hline
1000  &  $0\%$   &0.0101732  &&0.0134446&   \\
1000  &  $3\%$   &0.0102825 &&0.0135131&   \\
1000  &  $5\%$   &0.0104686  &&0.0136324&   \\
\hline\hline
1500  &  $0\%$   &0.00922353   &&0.012105&   \\
1500  &  $3\%$   &0.00937059   &&0.0121974&   \\
1500  &  $5\%$   &0.00922353   &&0.012357&   \\
\hline\hline
2500  &  $0\%$   &0.00815652   &&0.0105997&  \\
2500  &  $3\%$   &0.0083688   &&0.0107343&  \\
2500  &  $5\%$   &0.00870931   &&0.0109624&  \\
\hline\hline
\end{tabular}
\end{table}

\begin{table}
\caption{95\% C.L. bounds on the anomalous $\lambda_{tq\gamma}$ coupling for center of mass energy $\sqrt{s}=1.5$ TeV through the processes $ {\gamma}{\gamma}{\rightarrow}{ j }\bar{j}{ b} \bar{q} $ and ${\gamma} {\gamma}{\rightarrow}  \textit{l} {\nu}_\textit{l} b \bar{q}$ and .It is taken under consideration, systematic error as $0\%$,$3\%$,$5\%$ and integrated luminosities as 100,500,1000,1500,2500 $fb^{-1}$.}
\begin{tabular}{cccccc}
\hline
\multicolumn{6}{c}{$\sqrt{s}=1.5 $ TeV } \\
\hline
\multicolumn{2}{c}{} & \multicolumn{2}{c}{Hadronic Channel} & \multicolumn{2}{c}{Leptonic Channel} \\
\hline
${\cal L} \, (fb^{-1})$  & \hspace{0.5cm} $ \delta_{sys}$ \hspace{0.5cm}  &
\hspace{1.5cm} $\lambda_{q}$ \hspace{1.5cm} &&
\hspace{1.5cm} $\lambda_{q}$ \hspace{1.5cm} & \\
\hline\hline
100  &  $0\%$   &0.00480661   &&0.00669162&   \\
100  &  $3\%$   &0.00523605  &&0.00697708&   \\
100  &  $5\%$   &0.00580293   &&0.00740572&   \\
\hline\hline
500  &  $0\%$   & 0.00329334   &&0.00470136&   \\
500  &  $3\%$   & 0.00431143   &&0.00547035&   \\
500  &  $5\%$   & 0.00520801  &&0.0063018&   \\
\hline\hline
1000  &  $0\%$   &0.00280796  &&0.00406577&   \\
1000  &  $3\%$   &0.00413974   &&0.00514154&   \\
1000  &  $5\%$   &0.00511674   &&0.00610814&   \\
\hline\hline
1500  &  $0\%$   & 0.00256091  &&0.00374327&   \\
1500  &  $3\%$   & 0.00407713  &&0.00501392&   \\
1500  &  $5\%$   & 0.00508515   &&0.00603875&   \\
\hline\hline
2500  &  $0\%$   & 0.00228358   &&0.00338243&  \\
2500  &  $3\%$   & 0.00402475   &&0.00490308&  \\
2500  &  $5\%$   & 0.00505942   &&0.00598125&  \\
\hline\hline
\end{tabular}
\end{table}

\begin{table}
\caption{Same as in Table I, but for $\sqrt{s}=3.0 $ TeV.}
\begin{tabular}{cccccc}
\hline
\multicolumn{6}{c}{$\sqrt{s}=3.0 $ TeV} \\
\hline
\multicolumn{2}{c}{} & \multicolumn{2}{c}{Hadronic Channel} & \multicolumn{2}{c}{Leptonic Channel} \\
\hline
${\cal L} \, (fb^{-1})$  & \hspace{0.5cm} $ \delta_{sys}$ \hspace{0.5cm}  &
\hspace{1.5cm} $\lambda_{q}$ \hspace{1.5cm} &&
\hspace{1.5cm} $\lambda_{q}$ \hspace{1.5cm} & \\
\hline\hline
100  &  $0\%$   & 0.0139793   &&0.0187968&   \\
100  &  $3\%$   & 0.0140044   &&0.0188135&   \\
100  &  $5\%$   & 0.0140488   &&0.0188432&   \\
\hline\hline
1000  &  $0\%$   & 0.00802822   &&0.0103486&   \\
1000  &  $3\%$   & 0.00816624   &&0.0104416&   \\
1000  &  $5\%$   & 0.00839488   &&0.0106013&   \\
\hline\hline
3000  &  $0\%$   & 0.00619342   &&0.00774389&  \\
3000  &  $3\%$   & 0.00649248   &&0.0079506&  \\
3000  &  $5\%$   & 0.00693574   &&0.00828451&  \\
\hline\hline
4000  &  $0\%$   & 0.00579084   &&0.00717239&   \\
4000  &  $3\%$   & 0.00615332   &&0.00742576&   \\
4000  &  $5\%$   & 0.00666680   &&0.00782405&   \\
\hline\hline
5000  &  $0\%$   & 0.00549800   &&0.00675667&  \\
5000  &  $3\%$   & 0.00591705   &&0.00705268&  \\
5000  &  $5\%$   & 0.00648723   &&0.00750628&  \\
\hline\hline
\end{tabular}
\end{table}

\begin{table}
\caption{Same as in Table II, but for $\sqrt{s}=3.0$ TeV.}
\begin{tabular}{cccccc}
\hline
\multicolumn{6}{c}{$\sqrt{s}=3.0$ TeV } \\
\hline
\multicolumn{2}{c}{} & \multicolumn{2}{c}{Hadronic Channel} & \multicolumn{2}{c}{Leptonic Channel} \\
\hline
${\cal L} \, (fb^{-1})$  & \hspace{0.5cm} $ \delta_{sys}$ \hspace{0.5cm}  &
\hspace{1.5cm} $\lambda_{q}$ \hspace{1.5cm} &&
\hspace{1.5cm} $\lambda_{q}$ \hspace{1.5cm} & \\
\hline\hline
100  &  $0\%$   & 0.00444942   &&0.00492852&   \\
100  &  $3\%$   & 0.00467085   &&0.00506216&   \\
100  &  $5\%$   & 0.00500227   &&0.00527645&   \\
\hline\hline
1000  &  $0\%$   & 0.00244292   &&0.00277697&   \\
1000  &  $3\%$   & 0.00327872   &&0.00335318&   \\
1000  &  $5\%$   & 0.00402259   &&0.00395722&   \\
\hline\hline
3000  &  $0\%$   & 0.00182445   &&0.00211304&  \\
3000  &  $3\%$   & 0.00307643   &&0.00305432&  \\
3000  &  $5\%$   & 0.00391611   &&0.00378807&  \\
\hline\hline
4000  &  $0\%$   & 0.00168877   &&0.00196727&   \\
4000  &  $3\%$   & 0.00304829   &&0.00301013&   \\
4000  &  $5\%$   & 0.0039022    &&0.00376525&   \\
\hline\hline
5000  &  $0\%$   & 0.00159008   &&0.00186121&  \\
5000  &  $3\%$   & 0.00303104   &&0.00298266&  \\
5000  &  $5\%$   & 0.00389378   &&0.00375136&  \\
\hline\hline
\end{tabular}
\end{table}

 As it can be clearly seen in tables II and IV, this situation is particularly more evident when the anomalous coupling constant values derived from the CB photon interactions are examined. This is due to the fact that the cross section values of CB photon interactions are larger than the WW photon ones. In this case, increases in the cross section values lead to that the number of events (N) will be elevated, resulting in a reduction of the statistical error function $ {\delta}_{stat}$ and the observed changes in the systematic error function ${\delta}_{sys}$ to be significant. On the other hand, since CLIC has very high luminosity opportunities, statistical error function become smaller and thus, the differences in systematic error values become more noticeable on obtained limits for the anomalous coupling constant. To identify above mentioned effects of the both high luminosity and cross section values on anomalous couplings, we have analysed the variations of the branching ratio of anomalous $t\rightarrow q\gamma$ interactions via integrated luminosities for three different $ {\delta}_{sys}$ values for CB photons with hadronic channel. The obtained results are depicted in Figures 5 and 6.
  
 \begin{figure}
\includegraphics[width=10.cm]{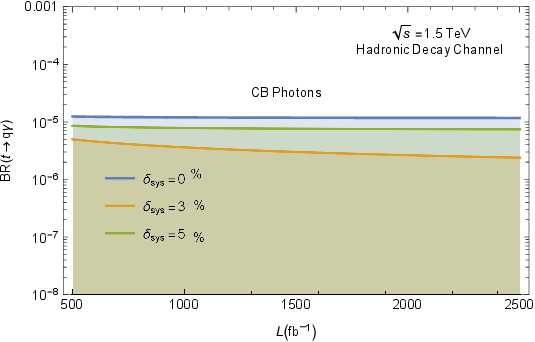}
\caption{For processes $\gamma \gamma {\rightarrow} j \bar{j}  b  \bar{q}   $ center-of mass energy of $\sqrt{s}=3.0TeV$ at the CLIC.}
\end{figure}
\begin{figure}
\includegraphics[width=10.cm]{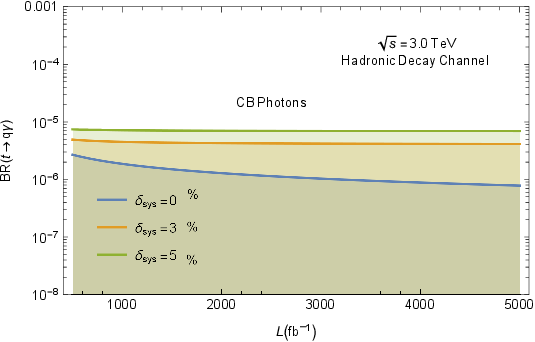}
\caption{Same as Fig5, but for center-of mass energy of $\sqrt{s}=3.0TeV$ at the CLIC.}
\end{figure}
  
  As it is seen in the figures, the constraints on the branching ratio is of the order $10^{-6}$ when the assumption ${\delta}_{sys}$ is set to $0\%$. In contrast, when $ {\delta}_{sys}$ is set to $5\%$, the constraints on the branching ratio worsen up to the order $10^{-5}$. Despite of the worsening trend, the obtained constraints are still the same order as the recent experimental constraints.
With this mind, in the following part of the study, to omit the effect of $ {\delta}_{sys}$ on the branching ratio, the integrated luminosity dependence of the branching ratio is examined by assuming ${\delta}_{sys}$ for handled all processes and center of mass energy values.

In this regard, in Figures 7 and 9, it is displayed branching ratio of anomalous $t\rightarrow q\gamma$ interactions as a function of integrated luminosities via ${\gamma \gamma}{\rightarrow} {\textit{l}} {\nu}_\textit{l}  b  \bar{q} $ ,  ${e^{+}} {e^{-}} {\rightarrow} {e^{+}} {\gamma}^{*} {\gamma}^{*} {e^{-}}{\rightarrow}  {e^{+}} \textit{l}{} {\nu}_\textit{l}  b  \bar{q}  {e^{-}} $  processes while  in Figures 8 and 10, $\gamma \gamma{\rightarrow} j \bar{j}  b  \bar{q}  $ and  ${e^{+}} {e^{-}} {\rightarrow} {e^{+}} {\gamma}^{*} {\gamma}^{*} {e^{-}}{e^{-}} {\rightarrow}  {e^{+}} j \bar{j} b \bar{q}  {e^{-}} $ processes are represented together. Here, center of mass energy are 1.5 TeV for Figures 7 and 9 and 3.0 TeV for Figures 8 and 10.

\begin{figure}
\includegraphics[width=10.cm]{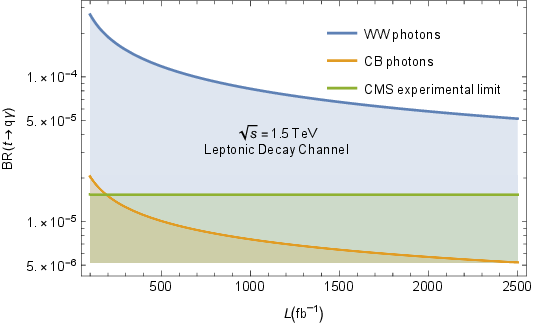}
\caption{For processes ${\gamma \gamma}{\rightarrow} {\textit{l}} {\nu}_\textit{l}  b  \bar{q} $ ,  ${e^{+}} {e^{-}} {\rightarrow} {e^{+}} {\gamma}^{*} {\gamma}^{*} {e^{-}} {\rightarrow}  {e^{+}} \textit{l}{} {\nu}_\textit{l}  b  \bar{q}  {e^{-}} $, 95\% C.L. sensitivity limits on  $BR(t \rightarrow q\gamma)$
for various integrated luminosities CLIC.}
\end{figure}
\begin{figure}
\includegraphics[width=10.cm]{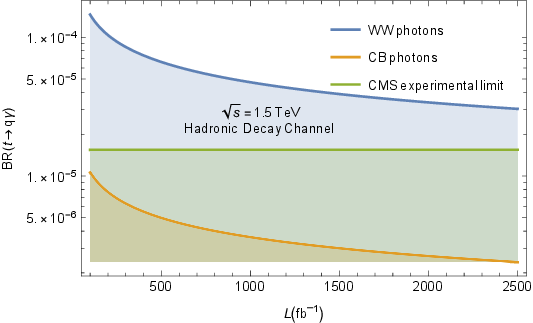}
\caption{For processes $\gamma \gamma {\rightarrow} j \bar{j}  b  \bar{q}   $ and  ${e^{+}} {e^{-}} \rightarrow {e^{+}} {\gamma}^{*} {\gamma}^{*} {e^{-}}  {\rightarrow}  {e^{+}} j \bar{j} b \bar{q}  {e^{-}} $, 95\% C.L. sensitivity limits on  $BR(t \rightarrow q\gamma)$
for various integrated luminosities CLIC.}
\end{figure}
\begin{figure}
\includegraphics[width=10.cm]{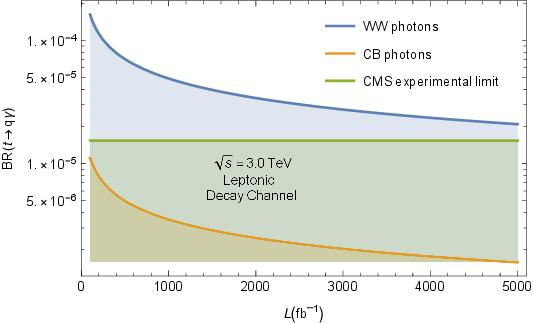}
\caption{Same as Fig7, but for center-of mass energy of $\sqrt{s}=3.0TeV$ at the CLIC.}
\end{figure}
\begin{figure}
\includegraphics[width=10.cm]{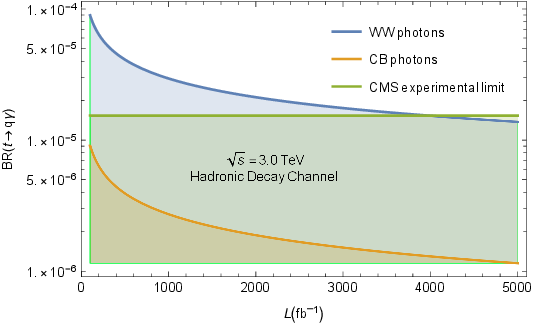}
\caption{Same as Fig7, but for center-of mass energy of $\sqrt{s}=3.0TeV$ at 
 the CLIC.}
\end{figure}

When Figures 7-10 are evaluated together, the most restrictive constraints on the branching ratio are obtained for CB photons in hadronic channel interactions for the center of mass energy of 3 TeV and the luminosity value of 5000$fb^{-1}$, as seen in the Figure 10. Here, the value of $BR(t\rightarrow q\gamma)$ is $1.18*10^{-6}$, which is 20 times more restrictive than results reported by the ATLAS and CMS experimental groups. Figure 10 have been examined for WW photon interactions at the same energy and luminosity value, it is seen that $BR(t\rightarrow q\gamma)$ is $1.31*10^{-5}$, which is slightly better than the current experimental constraints. Another remarkable point is that, as can be seen in Figure 8, even at the lower center of mass energy of 1.5 TeV investigated, the constraints obtained for CB photon interactions at a relatively low luminosity value, such as ${\cal L}=100fb^{-1}$, are more stringent than the experimental data. The $BR(t\rightarrow q\gamma)$ values obtained for the hadronic and leptonic channel at ${\cal L}=100fb^{-1}$ for the CB photons are $1.01*10^{-5}$ and $1.89*10^{-5}$, respectively. On the other hand, when the limit values obtained for the processes with WW photons are examined, it is seen that they are less restrictive than the experimental results except for a single exception, that is, they are above the experimental constraints. The only exception situation is valid for the high luminosity values ${\cal L}>3000fb^{-1}$ of the hadronic channel interactions at 3 TeV, where the limits obtained are found to be approximately at the same level as the experimental results. In this respect, making a quantitative comparison of the CB and WW photon results based on the data in the figures, it has been seen that CB photons provide significantly higher sensitivity. For instance, in calculations at a 1.5 TeV center-of-mass energy in the hadronic channel, the cross-section obtained with CB photons is nearly double that achieved with WW photons (as shown in Figures 1 and 2). This suggests that CB photons yield stronger signal strength and are more effective at suppressing background events. At the 3.0 TeV energy level, especially at high luminosity values (e.g.,  ${\cal L}=5000fb^{-1}$ ), the branching ratio limit for $BR(t\rightarrow q\gamma)$ obtained with CB photons is calculated to be $1.18*10^{-6}$, while for WW photons, the corresponding limit is approximately $1.31*10^{-5}$. This result shows that CB photons offer nearly 10 times greater sensitivity compared to WW photons under similar conditions. These values highlight that the narrower spectrum and higher cross-section values provided by CB photons optimize the signal-to-background ratio, thereby enhancing sensitivity. Additionally, the lower background level with CB photons makes them a superior option for new physics searches compared to WW photons. . Also, the impact of systematic uncertainties on sensitivity limits for FCNC $t\rightarrow q\gamma$ transitions can be quantitatively evaluated across various luminosities. At lower luminosities (e.g., $100 fb^{-1}$ ), systematic uncertainties, such as photon flux and cross-section variations, moderately influence sensitivity, resulting in deviations up to approximately 10\%. As luminosity increases (e.g., $1000fb^{-1}$ )), the effect of uncertainties becomes more pronounced, with deviations ranging from 15-25\% due to amplified photon flux and background modeling impacts. At high luminosities ($5000fb^{-1}$ )), systematic uncertainties in photon flux and detector efficiency significantly alter sensitivity limits, with potential deviations exceeding 25-30\%, especially in CB photon interactions. This analysis highlights the need for stringent control over systematic errors to optimize sensitivity in high-luminosity FCNC studies at CLIC. In summary, in the light of the obtained findings, CB photons advantages in sensitivity and background suppression make them a superior choice for investigating new physics, particularly in high-luminosity hadronic channel studies, where they offer the greatest potential for stringent FCNC constraints.

On the other hand, in our last reported work, the sensitivity limits on $BR(t\rightarrow q\gamma)$ were obtained by processes resulting from photon-proton interactions at the FCC-ep collider, and the most restrictive results were computed as $BR(t\rightarrow q\gamma)=2.01*10^{-6}$  through the ${e^{-}} {p} {\rightarrow} {e^{-}} {\gamma} {p} {\rightarrow} {e^{-}} W^{+} b {\gamma}{\rightarrow} {e^{-}} {j} {\bar{j}} b {\gamma}$ process at $\sqrt{s}=10.0 $ TeV and ${\cal L}=3000fb^{-1}$ values\cite{14}. 
In this presented study, the most restrictive $BR(t\rightarrow q\gamma)$ value was calculated as $1.18*10^{-6}$ for the $\gamma \gamma {\rightarrow} j \bar{j} b \bar{q} $ process, and it can be clearly said that this results obtained in the study is approximately 2 times more restrictive than the study we mentioned above and reported in the literature. Moreover, the photon-photon interaction processes in the current study have a cleaner environment independent of strong interactions compared to the photon-proton interactions in the study we previously reported in the literature, allowing to enable us that new physics contributions are more easily distinguishable from SM backgrounds. As a result, these findings may shed light on experimental studies on FCNC interactions to be conducted after the LHC.

\section{Conclusion}

In this study, anomalous FCNC transitions of the top quark, which is an important field of study in the new physics beyond the Standard Model, are examined. In these investigations, the existence of anomalous $t\rightarrow q\gamma$ interactions have been attempted to be revealed through the utilisation of diverse processes involving CB and WW photons. The calculations presented here are based on photon-photon interactions at CLIC, a future linear lepton collider. 
In this regard, the most stringent constraints on the branching ratio have been obtained with the 
$\gamma \gamma {\rightarrow} j \bar{j} b \bar{q} $ which is called the hadronic channel process formed via the interactions of CB photons, and in these calculations made for two different center of mass energies, the values of $BR(t\rightarrow q\gamma)$ have been found in the range of $9*10^{-6}-1*10^{-6}$ according to the varying luminosity values.
 On the other hand, in the calculations of CB photon interactions for the leptonic channel, the limit range of $BR(t\rightarrow q\gamma)$ is found to be $2*10^{-5}-2*10^{-6}$. These results for the CB photon are an order of magnitude more restrictive than the recent experimental limits reported by the CMS and ATLAS collaborations at the LHC. These findings suggests that the CB photon interaction processes in the study could be an attractive subject for the study of FCNC processes in new physics researchs. Furthermore, in the calculations performed via WW photons, the branching ratio of the $t\rightarrow q\gamma$ transitions calculated for hadronic channel interactions at $\sqrt{s}=3.0 $ TeV and ${\cal L}>3000fb^{-1}$ is found to be in the order of $1*10^{-5}$ and only the $BR(t\rightarrow q\gamma)$ values under these conditions are more restrictive than the current experimental results. Another advantage of the presented study is that the $\gamma\gamma$ and $\gamma^{*}\gamma^{*}$ processes investigated provide a clean background independent of proton remnants. This is one of the reasons why they are also promising for new physics research. In a brief, the all findings obtained in the study indicate that the processes considered may play a key role in the beyond-SM studies at the CLIC.

\end{document}